\documentclass[%
 aip,
 amsmath,amssymb,
 reprint,%
]{revtex4-1}

\usepackage{graphicx}
\usepackage{dcolumn}
\usepackage{bm}
\usepackage{hyperref}

\usepackage[utf8]{inputenc}
\usepackage[T1]{fontenc}
\usepackage{mathptmx}
\usepackage{etoolbox}
\usepackage{siunitx}
\usepackage{amsmath}  

\makeatletter
\def\@email#1#2{%
 \endgroup
 \patchcmd{\titleblock@produce}
  {\frontmatter@RRAPformat}
  {\frontmatter@RRAPformat{\produce@RRAP{*#1\href{mailto:#2}{#2}}}\frontmatter@RRAPformat}
  {}{}
}%
\makeatother
\begin{document}

\preprint{AIP/123-QED}

\title[Droplet impact on surfactant-laden thin liquid films]{Droplet impact on surfactant-laden thin liquid films: Vortex ring dynamics}
\author{Hatim Ennayar}
 \altaffiliation[\textbf{Author to whom correspondence should be addressed:} ennayar@sla.tu-darmstadt.de]{}
 \affiliation{ 
Institute for Fluid Mechanics and Aerodynamics, Technische Universität Darmstadt, Darmstadt, Germany \\ 
}
 \author{Hyoungsoo Kim}%
\affiliation{ 
Department of Mechanical Engineering, Korea Advanced Institute of Science and Technology, Daejeon \\ 
}%
\author{Jeanette Hussong}%
\affiliation{ 
Institute for Fluid Mechanics and Aerodynamics, Technische Universität Darmstadt, Darmstadt, Germany \\
}%


\date{\today}

\begin{abstract}
Droplet impact on surfactant-laden thin liquid films is investigated experimentally with emphasis on vortex ring dynamics. Bottom- and side-view imaging reveal that increasing surfactant concentration progressively stabilize vortex rings, suppress azimuthal instabilities and promote concentric mixing patterns. A regime map is established in terms of film thickness, Reynolds number, and surface-tension ratio, yielding an empirical instability threshold. Shadowgraphy observations suggest that Marangoni stresses modify early capillary-wave dynamics, potentially altering vortex ring formation and delaying instability onset. These findings clarify the link between interfacial stresses, vortex ring dynamics, and mixing patterns in thin-film droplet impact.
\end{abstract}

\maketitle

The impact of liquid droplet on a thin liquid film is a fundamental process encountered in a wide range of natural and industrial applications, such as rainfalls \cite{kinnell2005raindrop}, spray coating \cite{andrade2013drop,huang2018understanding}, inkjet printing\cite{wijshoff2018drop,lohse2022fundamental}, cosmetics \cite{srinivasan2015impact} and pharmaceutical \cite{christodoulou2018mathematical} applications, where controlled spreading and mixing is essential. In many of these situations the impacting droplet and the underlying liquid film differ in composition or contain surface-active agents, which results in non-uniform interfacial properties during impact. Such variations commonly produce surface tension gradients arising from uneven distributions of surfactants or contaminants. These gradients generate Marangoni stresses that drive tangential interfacial flow toward regions of higher surface tension, thereby influencing the impact dynamics. 

Previous investigations of surface-tension–driven effects in droplet impact have focused predominantly on impacts onto dry solid substrates or deep liquid pools \cite{kim2015spontaneous,kim2017solutal,jia2022three,pant2023marangoni,varghese2024effect}. Impacts on thin liquid films, on the other hand, have received comparatively less attention despite the different dynamics introduced by finite film thickness. Thin films are typically characterized using the non-dimensional thickness parameter $\delta=h/D$, defined as the ratio of the initial liquid film thickness $h$ to the droplet diameter $D$. Based on previous classifications, the thin-film regime is commonly associated with $\delta \leq 0.6$, where both wall proximity and finite film thickness influence the impact dynamics \cite{tropea1999impact,geppert2019experimental,liang2016review}. Within this regime, recent studies indicate that surfactants can substantially modify the impact outcome. \citet{constante2023impact} used fully three-dimensional direct numerical simulations to investigate the crown-splash regime and showed that surfactants significantly influence the late-stage dynamics by delaying ligaments breakup through Marangoni stresses. Complementary experimental and numerical work by \citet{quetzeri2025droplet} demonstrated that surfactant kinetics and transport modify crown morphology and splashing thresholds, with dynamic surface tension at crown-collapse times providing a robust predictor of impact outcome. In addition to modifying macroscopic impact dynamics, surfactants have been shown to strongly influence mixing patterns in thin liquid films. Using dye-based visualization, \citet{che2017impact} showed that droplet impact on surfactant-laden thin films leads predominantly to concentric mixing patterns, whereas flower-like structures are observed in surfactant-free films under comparable conditions. In explaining the absence of flower-shaped mixing patterns in surfactant-laden films, \citet{che2017impact} attributed the observed axisymmetric, concentric structures to surface rigidification induced by surfactants. In this context, rigidification refers to the retardation of interfacial dynamics caused by tangential Marangoni stresses that oppose inertia-driven interfacial flow, as demonstrated in by \citet{constante2021role}. However, existing studies neither identify the physical origin of the flower-shaped patterns observed in surfactant-free films, nor do they provide a mechanistic explanation for how surface rigidification reorganizes the subsurface flow to produce concentric mixing, leaving the link between interfacial stresses and bulk mixing dynamics unresolved. 

\begin{figure*}
\centering
\includegraphics{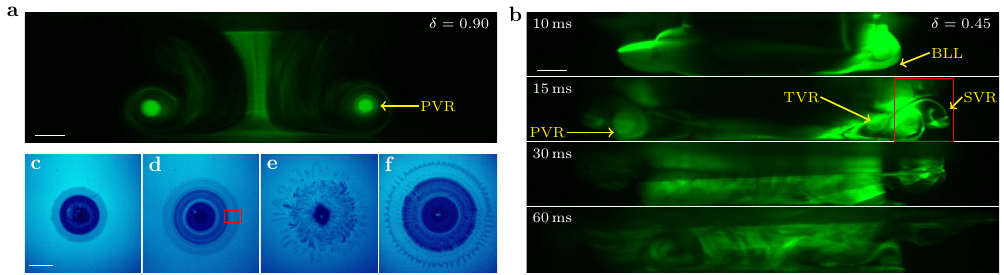}
  \caption{Vortex ring dynamics and associated mixing patterns during droplet impact on liquid films in the absence of surfactants. \textbf{(a)} For thick films ($\delta=0.90$), impact generates a primary vortex ring (PVR) that expands radially after interaction with the wall. Scale bar is equivalent to 1 mm. \textbf{(b)} For thin films ($\delta=0.45$), stronger wall interaction promotes boundary layer lift-off (BLL) and the formation of secondary (SVR) and tertiary (TVR) vortex rings, eventually leading to vortex rings breakdown. Scale bar is equivalent to 1 mm. \textbf{(c-f)} Bottom-view fluorescence visualizations showing the evolution of mixing patterns. Scale bar is equivalent to 3 mm. \textbf{(c)} expansion of primary vortex ring. \textbf{(d)} concentric ring structures corresponding to multiple vortex rings regime. \textbf{(e)} Vortex ring instability leading to breakdown. \textbf{(f)} Onset of azimuthal perturbations for thinner films ($\delta=0.09$). The red boxes highlight the region where vortex rings entrain droplet and film liquid in different proportions, producing concentration variations that appear in the bottom view as successive concentric mixing rings}
    \label{fig:1}
\end{figure*}

In a recent study \cite{ennayar2025vortex}, we investigated the baseline case of a droplet impacting a film of the same liquid in the absence of surfactants and demonstrated that the observed flower-like mixing structures originate from vortex ring dynamics generated during impact, whose evolution is influenced by wall interaction due to the finite film thickness. The vortex ring forms due to azimuthal vorticity production at the liquid–liquid interface caused by strong velocity gradients between the impacting droplet and the underlying liquid film \cite{cresswell1995drop}. Depending on the film thickness and impact conditions, three distinct phases were identified, of which only the first, the first two, or all three may occur, as illustrated in Fig. \ref{fig:1}. For thick films, or for impacts on thin films at lower velocities, the dynamics are governed by the radial expansion of the droplet-induced vortex ring following its interaction with the wall, as shown in Fig. \ref{fig:1}a. The associated mixing pattern observed from below remains axisymmetric at this stage, as illustrated in Fig. \ref{fig:1}c. As film thickness decreases, wall interaction becomes more pronounced. For example, at $\delta=0.45$ (Fig. \ref{fig:1}b), boundary layer separation at the wall triggers the second phase characterized by the formation of additional vortex rings. The regions highlighted in the red boxes (Figs. \ref{fig:1}b and d) show how the spiral motion the primary and secondary vortex rings entrains droplet and film liquid in different proportions, producing concentration variations that appear as successive concentric rings in the bottom view. At later times, azimuthal perturbations may develop, leading to the breakdown of the vortex rings, as shown in Fig. \ref{fig:1}e and Fig. \ref{fig:1}b at $t=\SI{30}{\milli\second}$ and $t=\SI{60}{\milli\second}$. For even thinner films ($\delta=0.09$), these azimuthal perturbations can be clearly seen in the associated mixing pattern on Fig. \ref{fig:1}f.

These findings establish vortex ring dynamics as a physical mechanism governing mixing in surfactant-free thin film impact. The question remains how this dynamics is altered when surface-active agents are present, particularly given earlier observations of a transition from chaotic flower-like to concentric mixing structures on surfactant-laden films \citet{che2017impact}. This letter addresses that gap by examining droplet impact on surfactant-laden thin liquid films with emphasis on vortex ring dynamics. For the first time, a systematic parametric investigation is conducted to investigate the influence of surfactants on the mixing outcome. The study focuses on the droplet deposition regime to avoid additional complexity associated with splashing and secondary droplet formation. Particular emphasis is placed on constructing a regime map for the onset of vortex ring instability, accounting also for surface tension differences between the droplet and the liquid film. 

\begin{figure}[htbp]
    \centering
    \includegraphics[scale=1]{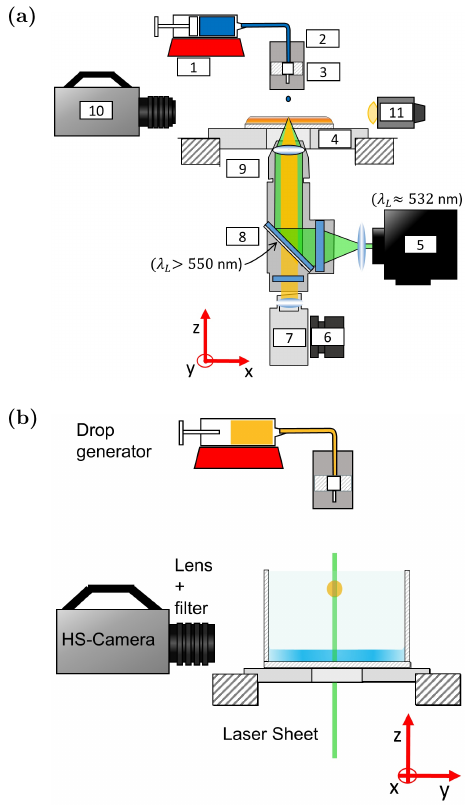}
       \centering
    \caption{\textbf{(a)} Schematic illustration of bottom-view experimental setup, together with a side-view shadowgraphy configuration for visualizing the interface evolution. (1) Syringe pump, (2) z-Traverse, (3) Cannula, (4) Liquid film on FTO glass substrate, (5) High power LED, (6) x,y,z-Traverse, (7) HS-Camera, (8) Dichroic mirror ($\lambda_L>\SI{550}{\nano\meter}$) with bandpass filters (absorption \textbf{$\SI{532}{\nano\meter}$} and emission \textbf{$\SI{575}{\nano\meter}$}), (9) Microscope Objective, 10) HS-Camera + lens, 11) LED. \textbf{(b)} Schematic representation of the side-view experimental configuration used for investigating vortex-ring dynamics illustrating the droplet generator, laser sheet illumination (\textbf{$\lambda_L=\SI{532}{\nano\meter}$}) and high-speed camera with Lens and Longpass filter ($\lambda_L>\SI{560}{\nano\meter}$).}
    \label{fig:2}
\end{figure}

The experimental configuration follows our previous work \cite{ennayar2025vortex} and is outlined here with emphasis on the aspects relevant to the present investigation. Droplet impact on thin liquid films was studied using complementary side- and bottom-view optical arrangements (see Fig. \ref{fig:2}) to characterize subsurface vortex ring dynamics and the resulting mixing patterns. Droplets were generated using a $\SI{5}{\milli\litre}$ syringe (Braun GmbH) driven by a precision syringe pump (Aladdin AL-1010, WPI). The syringe was connected via tubing to a blunt needle mounted on a motorized traverse, allowing precise adjustment of release height and impact velocity. Impact conditions were selected to reach Reynolds $Re=(UD)/\nu$ and Weber $We=(\rho U^2 D)/\sigma$ numbers up to $3300$ and $64$, while remaining in deposition regime. Here, $U$ denotes the impact velocity, $\nu$ the kinematic viscosity, $\rho$ the density and $\sigma$ the surface tension.

For bottom-view measurements, liquid films were prepared on \SI{50}{\milli\meter} $\times$ \SI{50}{\milli\meter} glass substrate (Sigma-Aldrich). Film thickness $h$ was adjusted using a syringe and measured with a chromatic-confocal point sensor with accuracy $\pm \SI{0.4}{\micro\meter}$ (confocalDT IFS2407-0.8, Micro-Epsilon), yielding non-dimensional thicknesses $\delta$ between $0.09$ and $0.45$. Mixing was visualized by laser-induced fluorescence using Rhodamine 6G added to the film, while droplets remained dye-free. Imaging was performed using a custom-built microscope \cite{ennayar2023lif,brockmann2022utilizing,brockmann2025enhancement} with an optical tube (InfiniTube Special, Infinity Photo-Optical), illuminated by a $\SI{7}{\watt}$ high-power green LED ($\lambda_L \approx \SI{532}{\nano\meter}$, ILA iLA.LPS v3) and appropriate dichroic and band-pass filters. High speed recording were obtained with a Phantom T1340 CMOS camera (Vision Research). Additionally, to resolve the early-time interface dynamics immediately after impact, shadowgraphy measurements were performed. The shadography system employed a Phantom T3610 high-speed camera operated at $22000~\mathrm{fps}$ and equiped with a Laowa 25 mm 2.5$\times$ lens. The impact region was back-illuminated using a high-intensity LED source (Veritas Constellation 120E).

Side-view measurements for vortex ring visualizations used the same droplet generator and a transparent acrylic cuvette (\SI{50}{\milli\meter} $\times$ \SI{50}{\milli\meter} $\times$ \SI{50}{\milli\meter}) treated to enhance wettability. A thin laser sheet ($\SI{532}{\nano\meter}$, Microvec) illuminated the impact plane, and vortex-ring dynamics were recorded using a Photron Fastcam Mini-AX200 high-speed camera. In this configuration, droplets were dyed with Rhodamine 6G instead of the liquid film, which does not affect water surface tension at the employed concentration \cite{seno2001}. To investigate surfactant effects, aqueous sodium dodecyl sulfate (SDS) solutions were prepared at controlled fractions of the critical micelle concentration (CMC$=8.2\,\mathrm{mM}$) to produce films with varied surface tension. For comparison between cases, a dimensionless surface tension ratio $\sigma^*=\sigma_f/\sigma_d$ is introduced, defined as the ratio of film surface tension to that of the impacting water droplet. The corresponding surfactant concentrations and surface tension values are summarized in Table \ref{tab:1}

\begin{figure*}[htbp]
    \centering
    \includegraphics[scale=1.0]{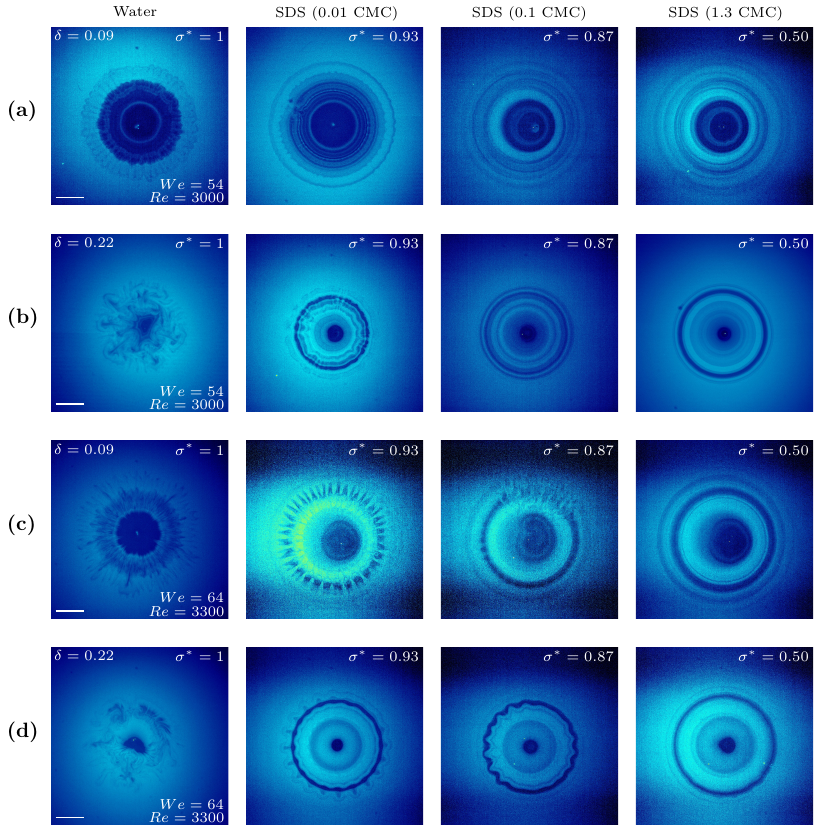}
       \centering
    \caption{Influence of surfactant concentration on mixing patterns following droplet impact on thin liquid films. Bottom-view fluorescence visualizations are shown for water droplets impacting aqueous films of pure water $\sigma^*=1$ and sodium dodecyl sulfate (SDS) solutions at different concentrations corresponding to $\sigma^*=0.93$ (0.01 CMC), $\sigma^*=0.87$ (0.1 CMC) and $\sigma^*=0.5$ (1.3 CMC). Cases are presented for different film thicknesses and impact conditions. \textbf{(a)} $\delta=0.09$, $Re=3000$ and $We=54$. \textbf{(b)} $\delta=0.22$, $Re=3000$ and $We=54$. \textbf{(c)} $\delta=0.09$, $Re=3300$ and $We=64$. \textbf{(d)} $\delta=0.22$, $Re=3300$ and $We=64$. Scale bar is equivalent to 2 mm.}
    \label{fig:3}
\end{figure*}

\begin{table}
\caption{\label{tab:1}Surface tension of the liquid films investigated, including water and aqueous SDS solutions prepared at different fractions of the critical micelle concentration (CMC). The droplets consist of water with density $\rho = \SI{997}{\kilogram\per\meter^3}$ and kinematic viscosity $\nu = \SI{1.004}{\milli\meter^2\per\second}$ at room temperature.}
\begin{ruledtabular}
\begin{tabular}{lcc}
\textbf{Solution in thin film} &
		\textbf{$\sigma_f$ (\si{\milli\newton\per\meter})} &
		\textbf{$\sigma^* = \sigma_f / \sigma_d$} \\
\hline
	Water                & 72.3 & 1.00 \\
		SDS (0.005 CMC)      & 68.7 & 0.95 \\
		SDS (0.01 CMC)       & 67.2 & 0.93 \\
		SDS (0.1 CMC)        & 62.7 & 0.87 \\
		SDS (0.2 CMC)        & 55.4 & 0.77 \\
		SDS (1.3 CMC)        & 36.2 & 0.50 \\
\end{tabular}
\end{ruledtabular}
\end{table}

Fig. \ref{fig:3} presents the influence of surfactant-induced surface tension gradient on the mixing patterns observed after droplet impact on thin liquid films for representative combinations of $\delta$, $Re$ and $We$. For $\delta=0.09$ at $Re=3000$, $We=54$, the presence of surfactants strongly alters the instability of the vortex ring compared to the reference water case. As shown in Fig.~\ref{fig:3}a, the pronounced azimuthal perturbations visible in the pure water film ($\sigma^* = 1$) are progressively weakened as the surfactant concentration increases and $\sigma^*$ decreases, and the mixing toward concentric ring structures. A similar trend is observed for the thicker film case ($\delta=0.22$) at identical impact conditions. In the absence of surfactant, vortex rings undergo rapid destabilization and break down into chaotic mixing structures as seen in Fig.~\ref{fig:3}b. Introducing even small amounts of surfactant, reducing the surface tension ratio $\sigma^*=0.93$, already suppresses this breakdown, although azimuthal instabilities remain visible. With further increase in SDS concentration, these instabilities diminish, and for $\sigma^*=0.87$, corresponding to a SDS concentration of $0.1$ CMC, the mixing pattern is characterized by well-defined concentric ring structures.

Increasing the impact energy further modifies this behavior. While at $Re=3000$ the onset of concentric mixing patterns was observed already at $\sigma^*=0.87$, higher impact velocities ($Re=3300$) reveal that azimuthal instabilities can persist even at this surface tension ratio. As shown in Figs. \ref{fig:3}c and d, localized azimuthal perturbations remains visible along portions of the vortex ring, indicating that increased inertia partially counteracts the stabilizing influence of Marangoni stresses. To quantify this effect, the instability boundary established for surfactant-free films in our previous work \cite{ennayar2025vortex} was revisited. In that study, the critical Reynolds number above which vortex ring instabilities emerge was described by
\begin{eqnarray}
Re = 1600 + 1500\,\delta,
\label{Eq:1}
\end{eqnarray}
for $\sigma^*=1$. The present measurements show that this critical boundary shifts systematically with increasing surfactant concentration. Increasing SDS concentration progressively delays the onset of instability. From concentrations around $0.2$ CMC ($\sigma^*=0.77$) and above, no vortex ring instability was observed within the explored range of $Re$, $We$ and $\delta$.

\begin{figure}
\centering
\includegraphics{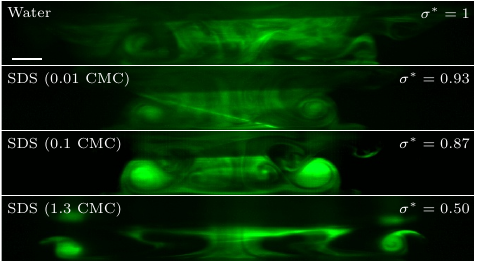}
  \caption{Side-vie LIF visualizations at $t=\SI{60}{\milli\second}$ illustrating the influence of surfactant concentration on vortex ring dynamics following droplet impact on thin liquid films ($\delta=0.45$, $Re=3000$, $We=54$). A water droplet impacts aqueous films of pure water $\sigma^*=1$ and sodium dodecyl sulfate (SDS) solutions at increasing concentrations corresponding to $\sigma^*=0.93$ (0.01 CMC), $\sigma^*=0.87$ (0.1 CMC) and $\sigma^*=0.5$ (1.3 CMC). Scale bar is equivalent to 1 mm.}
    \label{fig:4}
\end{figure}

Cross-sectional fluorescence images from a side-view provide further insights how vortex ring structures alter with increasing surface tension difference. Fig. \ref{fig:4} shows representative cases for liquid films of thickness $\delta=0.45$ with varying surfactant concentrations, recorded at $t=\SI{60}{\milli\second}$ after droplet impact for $Re=3000$ and $We=31$. In the reference case of pure water ($\sigma^*=1$), the decay of the vortex rings into chaotic mixing structures is clearly observed. This behavior follows the successive formation of multiple vortex rings induced by boundary layer separation, as previously shown in Fig.~\ref{fig:1}b, and their subsequent structural breakdown. With the introduction of surfactants, the vortical structures remain noticeably more coherent during their interaction with the wall as the SDS concentration rises, indicating a stabilization of the vortex dynamics, as also shown in the supplementary videos.

\begin{figure}
\centering
\includegraphics[scale=1]{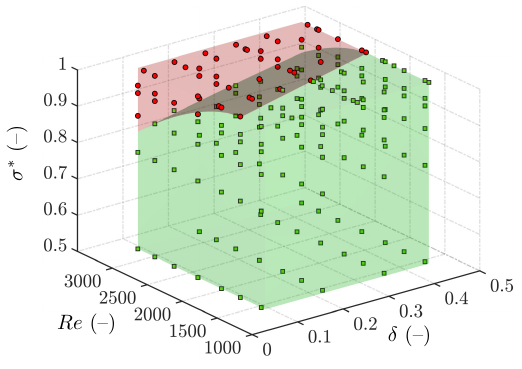}
  \caption{Regime map for the onset of azimuthal vortex ring instability in terms of dimensionless film thickness, Reynolds number and surface tension ratio ($\delta,Re,\sigma^*$). Red markers denote cases exhibiting vortex ring instability, whereas green markers correspond to stable, axisymmetric vortex evolution. The shaded surface represents the instability boundary inferred from a non-linear support vector machine classifier, while the colored regions indicate stable (green) and unstable (red) regimes. The boundary shifts toward higher Reynolds numbers with decreasing $\sigma^*$, reflecting the progressive stabilization by surfactant-induced Marangoni stresses.}
    \label{fig:5}
\end{figure}

To characterize the onset of vortex ring instability in the presence of surfactants, a regime map was constructed as a function of dimensionless film thickness $\delta$, Reynolds number $Re$ and surface tension ratio $\sigma^*$ as illustrated in Fig. \ref{fig:5}. Experimental cases exhibiting azimuthal perturbations with or without vortex ring breakdown ar marked in red, while cases with concentric ring structures are marked in green. A non-linear support vector machine \cite{cortes1995support} classifier with a Gaussian kernel was employed to infer the continuous boundary separating these regimes. Building on the instability criterion previously established for surfactant-free films expressed in Eq. (\ref{Eq:1}), the classifier-derived boundary was further approximated by introducing a nonlinear dependence on the surface tension ratio. This yield an empirical threshold of the form
\begin{eqnarray}
Re_{c}(\delta,\sigma^{*}) = 1600 + 1500\,\delta + A\,(1-\sigma^{*})^{p},
\label{Eq:2}
\end{eqnarray}
where $A=4348$ and $p=0.62$ are empirical fitting parameters. The resulting surface, shown in Fig. \ref{fig:5}, reveals a shift of the instability threshold towards higher Reynolds numbers as $\sigma^*$ decreases, indicating progressive stabilization of vortex ring dynamics due to the influence of surfactant-induced Marangoni stresses.

To better understand the observed stabilization of vortex ring with decreasing surface tension ratio, it is useful to consider the physical origin of the instability itself. In our previous study \cite{ennayar2025vortex}, the onset of instability was found to correlate strongly with vortex ring strength, quantified through its circulation. Weak vortex rings generated at low inertia impacts remained stable after interacting with the wall, whereas stronger rings with higher circulation were developed instabilities following wall interaction. This suggests that vortex circulation constitutes a key parameter governing the transition to instability.

\begin{figure*}
\centering
\includegraphics[scale=1]{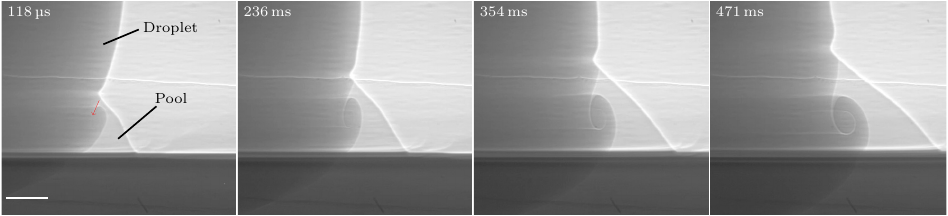}
  \caption{Time-resolved X-ray visualizations of droplet impact on a pool of same liquid, illustrating the mechanism of vortex ring generation at the liquid-liquid interface \cite{lee2015origin}. The red arrow highlights the region where the localized penetration si associated with the coiling of vortical structures. Scale bar is equivalent to $\SI{150}{\micro\meter}$. Images adapted from supplementary videos of \citet{lee2015origin}. licensed under \href{https://creativecommons.org/licenses/by/4.0/}{CC BY 4.0}.}
    \label{fig:6}
\end{figure*}

In this context, the present results motivate our hypothesis that surfactant-induced Marangoni stresses modify early-time generation of vorticity during impact, thereby influencing the strength of the resulting vortex ring and consequently the onset of subsequent instability. A possible mechanism can be inferred from the work of \citet{lee2015origin}, who identified capillary waves as a primary source of azimuthal vorticity during droplet impact. They showed that once capillary waves propagate over one wavelength, momentum transport associated with these waves induces localized penetration of the pool fluid into the droplet, accompanied by the formation and coiling of vortical structures. This process is illustrated in Fig. \ref{fig:6}, reproduced from \citet{lee2015origin}, where X-ray imaging reveals the penetration and the associated vortex coiling (red arrow).

Recent studies indicate that surfactant-induced Marangoni stresses can substantially modify capillary-wave dynamics at fluid interfaces. Surfactant spreading generates surface tension gradients that produce coupled Marangoni ridges and capillary wave packets \cite{sauleda2022surfactant}. More recent work \cite{shi2025complex}, shows that localized surfactant deposition produces complex wave-packet evolution in which portions of the interface directly influenced by Marangoni stresses exhibit altered propagation characteristics compared with clean capillary waves, including wave splitting. These findings indicate that capillary-wave dynamics in surfactant-laden films can deviate substantially from the classical clean-interface behavior.

\begin{figure*}
\centering
\includegraphics[scale=1.05]{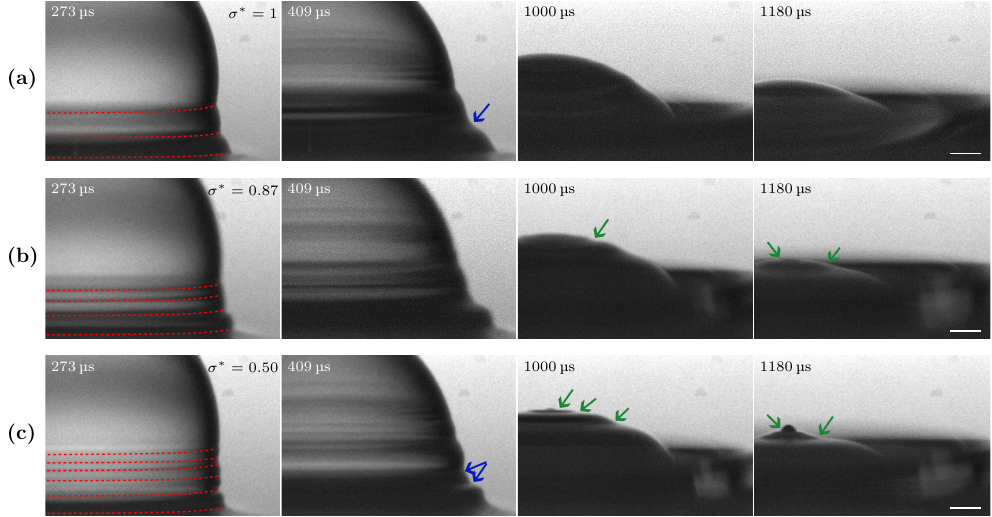}
  \caption{Time-resolved shadowgraphy images illustrating early interface dynamics during droplet impact on thin liquid films of thickness $\delta=0.22$ at $Re=3000$ and $We=54$ with different surfactant concentrations. \textbf{(a)} pure water ($\sigma^*=1$), \textbf{(b)} SDS solution at 0.1 CMC ($\sigma^*=0.87$) and \textbf{(c)} SDS solution at 1.3 CMC ($\sigma^*=0.50$). The red dashed lines indicate the ripples developed shortly after impact. The interface near the engulfment region, where the vortex ring formation initiates, becomes visibly altered with increased surfactant concentrations, as shown in blue arrows. For surfactant-laden films, surface ripples propagating toward the droplet apex are observed at $t \approx\SI{1}{\milli\second}$ (green arrows). Scale bar is equivalent to $\SI{500}{\micro\meter}$.}
    \label{fig:7}
\end{figure*}

Consistent with these observations, side-view imaging of droplet impact on surfactant-laden thin films (Fig. \ref{fig:7}) reveals clear modifications of interfacial wave dynamics. Increasing surfactant concentration leads to the appearance of multiple wave crests shortly after impact, consistent with wave-packet splitting and enhanced interfacial deformation. Concurrently, the interface near the engulfment region, where the vortex ring formation initiates, becomes visibly altered as shown in blue arrows. This is consistent with a possible modification of the penetration dynamics. In addition, surface ripples propagating toward the droplet apex are observed at approximately $\SI{1}{\milli\second}$ after impact, as shown in green arrows in Fig. \ref{fig:7}. These ripples are detected for surfactant-laden films and become more pronounced with increasing surfactant concentration, suggesting an influence of surfactant-induced interfacial stresses on wave propagation.

Based on these observations, surfactant-induced surface tension gradients appear to modify early-time capillary-wave propagation during impact. A plausible mechanism is that the associated Marangoni stresses redistribute momentum into tangential surface flows, thereby reducing the effective normal penetration responsible for vortex ring formation. This mechanism provides a possible explanation for the progressive weakening of vortex circulation and the delayed onset of azimuthal instability observed as the surface tension ratio decreases.

To our best knowledge, this work provides the first systematic experimental investigation of surfactant effects of droplet impact onto thin liquid films in this parameter regime, introducing a regime map that organizes the observed dynamics in terms of the relevant dimensionless parameters. Nevertheless, further subsurface-resolved investigations, such as X-ray imaging or fully coupled numerical simulations, would be valuable to directly quantify the influence of surfactants on vortex coiling and associated vorticity generation.

\begin{acknowledgments}
This project is funded by the Deutsche Forschungsgemeinschaft (DFG, German Research Foundation) – project number 237267381 – TRR 150, sub-project A07. \\
\end{acknowledgments}

\section*{Data Availability Statement}

The data that support the findings of this study are openly available in Zenodo at [add Zenodo reference].

\nocite{*}
\bibliography{aipsamp}

\end{document}